\documentclass[useAMS,usenatbib]{mn2e}
\usepackage{graphicx}

\title[Software design for panoramic astronomical pipeline processing]{Software design for panoramic astronomical pipeline processing}

\author[L. Shamir, R. J. Nemiroff, D. O. Torrey and W. E. Pereira]{L. Shamir$^{1}$\thanks{E-mail:lshamir@mtu.edu}, R. J. Nemiroff$^{2}$, D. O. Torrey$^{3}$ and W. E. Pereira$^{4}$ \\
$^{1}$Department of Computer Science, Michigan Technological University, 1400 Townsend Dr., Houghton, MI, 49931, USA \\
$^{2}$Department of Physics, Michigan Technological University, 1400 Townsend Dr., Houghton, MI, 49931, USA \\
$^{3}$East Engineering Computer Network, Michigan Technological University, 1400 Townsend Drive, Houghton, MI,  49931, USA \\
$^{4}$ThermoAnalytics, Inc., Calumet, MI 49913, USA}

\begin{document}

\date{}

\pagerange{\pageref{firstpage}--\pageref{lastpage}} \pubyear{2005}

\maketitle

\label{firstpage}

\begin{abstract}
We describe the software requirement and design specifications for all-sky panoramic astronomical pipelines. The described software aims to meet the specific needs of super-wide angle optics, and includes cosmic-ray hit rejection, image compression, star recognition, sky opacity analysis, transient detection and a web server allowing access to real-time and archived data. The presented software is being regularly used for the pipeline processing of 11 all-sky cameras located in some of the world's premier observatories. We encourage all-sky camera operators to use our software and/or our hosting services and become part of the global {\it Night Sky Live} network.
\end{abstract}


\maketitle

\section{Introduction}
In the past several years, all-sky cameras have become increasingly popular in the field of astronomy, and many systems incorporating super-wide fish-eye optics have been developed, installed and operated. Applications of all-sky cameras include cloud detection \citep{Hul94,Hog01,Tak03}, bright star monitoring \citep{Sha04,Gel04}, transient detection \citep{Kro01,Sha05e} and meteor science \citep{Sop92,Obe98,Bro02,Spu03}. However, while a substantial portion of the effort is directed toward the design of the hardware, less has yet been reported on the development of the software required for the autonomous operation of all-sky cameras. This software development can sometimes turn into an unexpectedly demanding task \citep{Bro87} and may delay the beginning of the effective operation of the camera.

In this paper we describe a software system that enables the operation and the analysis of astronomical pipelines. The algorithms, software and hosting services described in the paper can be used by all-sky camera operators who seek to increase the effectiveness of the data produced by their cameras, yet prefer to avoid the costs of software development. In Section~\ref{client_side} we describe the design of the software running on the remote station, including cosmic-ray hit rejection, star recognition, opacity maps and transient detection, and in Section~\ref{server_side} we describe the software design of the main server, which collects and archives the data transmitted from the remote stations.

\section{Remote station software design}
\label{client_side}

The software that runs on the remote station is based on Linux Red-Hat, which is considered robust comparing to other existing PC operating systems. The remote station starts taking exposures in time intervals of 236 seconds. For instance, if the duration of the exposure is set to 180 seconds, an image will be taken every 236 seconds. If the duration is set to 360 seconds, an image will be taken every 472 seconds. Since 236 second intervals are used, a given image can be compared with another image taken on a different day, but at the exact same sidereal time so that the positions of the stars appearing in the two images are identical. For instance, the positions of stars recorded by an image taken at 12:00:00 UT are the same as the positions of stars recorded on the previous day at 12:03:56 UT or at 12:07:52 two days before. This feature is used for all-sky opacity maps and transient detection, which will be described later in the paper.

Images are taken only after sunset (based on USNO data) and before dawn. No images are taken during the day. The system also changes the duration of the exposure according to the position and phase of the moon. When the moon is up, the duration of the exposure is shorter.

In order to support as many stations as possible, yet without overloading the main server, all CPU-intensive image processing is done on the computer hosting the camera (the remote station). This approach takes advantage of the free CPU resources available while the camera is taking an exposure.


\subsection{Cosmic ray-hit rejection}
\label{cosmic_ray_hits}

After each new exposure is taken, cosmic ray-hits are rejected from the frame so the image can be processed. One technique of cosmic-ray hit rejection is by comparing several exposures of the same field \citep{Win94,Fix00}. However, since one of the main purposes of the described system is to automatically detect optical transients that appear in single images \citep{Sha05e} as well as meteors \citep{Sha05b2}, such techniques are not suitable for this system. In order to reject cosmic-ray hits without rejecting true transients, a fuzzy-logic based algorithm for cosmic-ray hit rejection from single images \citep{Sha05a2} is applied. Like some other cosmic-ray hit rejection algorithms from single images, the rejections make use of the non-point source nature of cosmic ray splashes. Even though this algorithm does not introduce any better accuracy than other reported cosmic-ray hit rejection algorithms, it has a clear advantage in terms of computational complexity, which is a major concern in high image-rate pipelines. While some of the previously proposed algorithms may be too slow for systems at this scale \citep{Rho00,Van01,Pyc04}, the fuzzy-logic based algorithm processes a 1024$\times$1024 frame in just 12 seconds \citep{Sha05a2}.

\subsection{Star recognition}
\label{star_recognition}

In order to monitor bright stars, the software associates stellar objects visible in the frames with known-catalogued objects. This is done by applying a fuzzy-logic based star recognition algorithm designed specifically for super-wide angle astronomical images \citep{Sha05a}.

For each recognized star, the intensity of the local background is estimated and photometric data are collected. The intensity of the local background is estimated by the median of the 1600 pixels surrounding the point spread function (PSF). The intensity of the star is estimated by averaging the brightest pixels of the star's PSF and then subtracting the estimated local background. For each bright star (brighter than a certain preset magnitude), the averages of the brightest 1, 5, 9, 16 and 25 pixels are collected. For dimmer stars, however, the realized PSF is smaller so only the average of the top five pixels is currently listed. The average of the brightest pixels corresponds to their sum, but provides smaller numbers that can be handled using 16-bit integers. This reduces the bandwidth and storage space required to store these data.

Photometry files also list for each detected object some additional data taken from the Henry Draper catalogue such as the visual magnitude, the spectral type and celestial coordinates. The data are then stored in text-based HTML files that can be viewed by internet browsers and electronic spread sheets. The HTML files are xml-tagged, so that customized external software tools can easily browse and collect the data. The photometric precision of a single measurement is around 0.2 magnitudes, which allows star variability monitoring of high-amplitude variable stars such as Algol \citep{Muz05}. However, using measurements obtained over several nights can lead to photometric precision of 0.02 magnitudes, which allows monitoring low-amplitude variable stars such as Polaris \citep{Nem05a}.

Another product of this processing is the generation of annotated JPG images, which labels the names of the constellations and bright stars \citep{Sha04}. These images can be used for ``cosmetic'' purposes, but can also be applied to educational purposes such as studying the night sky \citep{Nem05b}.

\subsection{Opacity maps}
\label{opacity_maps}

All-sky cameras can be used as simple cloud monitoring devices by simply inspecting the images by eye \citep{Rut03,Per05,Per06}. However, while thick clouds can be easily noticed, thin cirrus clouds at night are often invisible to the unarmed human eye. In order to provide a more informative view of the opacity of the night sky, photometric measurements of numerous background stars are combined with simultaneous sky brightness measurements to differentiate thin clouds from sky glow sources such as air glow and zodiacal light \citep{Sha05b,Nem03}. This is done by comparing all-sky images to canonical images taken on other nights at the same sidereal time. The canonical images are collected automatically by counting the number of stars brighter than a certain magnitude detected by the star detection algorithm \citep{Sha05a}. If the PSFs of a sufficient percentage of expected stars are found in the image, the image is classified as {\it clear}, and added to the database of canonical images where it is co-added to all other clear images taken at the same sidereal time.

When an image is taken, any PSF brighter than its background by a given $\sigma$ is assigned with two values: a stellar intensity, and an estimated intensity of the background, as explained in Section~\ref{star_recognition}. The values assigned to the PSFs are then compared to the values of the PSFs of a canonical image. The comparison is based on Equations~\ref{star_glow} and~\ref{sky_glow}.

\begin{equation}
\label{star_glow}
M_{\mathrm star} = T \cdot I_{\mathrm star} + T \cdot I_{\mathrm space} + I_{\mathrm cloud}
\end{equation}

\begin{equation}
\label{sky_glow}
M_{\mathrm background}= T \cdot I_{\mathrm space} + I_{\mathrm cloud}
\end{equation}

where $M_{\mathrm star}$ is the intensity of the light coming from pixels at the location of a star in the given image, $M_{\mathrm background}$ is the intensity of the light coming from pixels {\it not} at the location of a star, $I_{\mathrm star}$ is the intensity of the star, $I_{\mathrm space}$ is the intensity of light from background space, $I_{cloud}$ is the intensity of the cloud covering the star and $T$ is the transmission around the star. Using Equations~\ref{star_glow} and~\ref{sky_glow}, the transmission is determined by Equation~\ref{sky_opacity}.

\begin{equation}
\label{sky_opacity}
T=\frac{(M_{\mathrm star} - M_{\mathrm background})}{I_{\mathrm star}} 
\end{equation}

where $I_{\mathrm star}$ is based on measurements taken from the canonical image such that $I_{\mathrm star}=Mo_{\mathrm star}-Mo_{\mathrm background}$, where $Mo_{star}$ and $Mo_{background}$ are the measured intensity of the star and the background, respectively, in the canonical image. This calculation of $I_{\mathrm star}$ is based on the assumption that no light is lost as starlight travels through air on a clear night. This assumption is not true, however, and provides a simplification of the problem that allows estimating the normalized transmission comparing to a clear night. This simplification gives 

\begin{equation}
\label{sky_opacity1}
T=\frac{(M_{\mathrm star} - M_{\mathrm background})}{(Mo_{\mathrm star} - Mo_{\mathrm background})} 
\end{equation}

Since the desired product is a broad relative transmission map, the transmission of each pixel is calculated by interpolating the transmission measured directly along the lines to the stars in the image.  This is performed by choosing, for example, the four nearest stars $S_{left}, S_{top}, S_{right}, S_{bottom}$ such that $|Y_{Sl}-Y_0|<X_0-X_{Sl}$, $|Y_{Sr}-Y_0|<X_{Sr}-X_0$, $|X_{St}-X_0|<Y_0-Y_{St}$, $|X_{Sb}-X_0|<Y_{Sb}-Y_0$, where $X_{Sn}$ is the X image coordinate of the star $S_n$, $Y_{Sn}$ is the Y image coordinate of the star $S_n$, and $(X_0,Y_0)$ are the image coordinates of the given pixel whose sky transmission is being estimated.  After finding the four nearest stars, a two-dimensional linear interpolation of the transmission measured by the four stars is performed, and the calculated value is determined as the relative transmission of the pixel. The computed relative transmission values of all pixels in the frame are then used for generating the all-sky opacity maps such that a high relative transmission implies high relative sky opacity and low relative transmission implies low relative opacity.

The computed normalized transmission $T$ provided by this method is not intended to be used for scientific purposes such as photometric reduction. However, providing a numeric indication of the estimated atmospheric transmission is believed to have an advantage over a simple boolean clear/cloudy bit of information. The numeric indication can be used by robotic telescopes for the purpose of decision making (for instance, weighing the importance of the observation against the atmospheric transmission) and can also provide human observers with more informative indications regarding the visibility conditions of the area of the sky they observe.

\begin{figure*}[ht]
\includegraphics{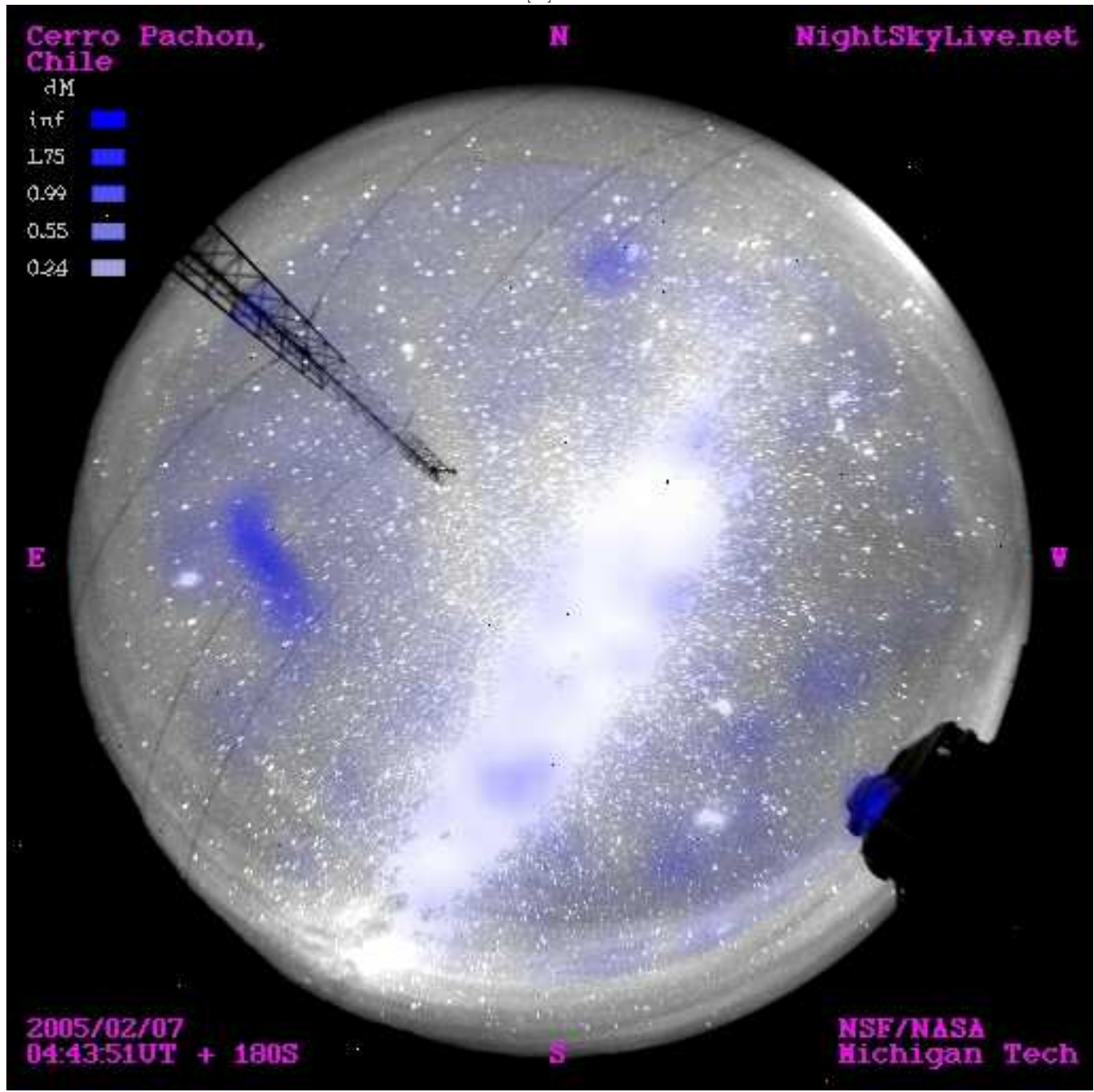}
\caption{All-sky opacity map generated in Cerro Pachon, Chile}
\label{opacity_map}
\end{figure*}

In order to visualize the sky opacity map, the pixel is coloured in blue such that a stronger blue colour represents higher opacity. The maps are generated by changing the B component of each RGB triple such that $B=B_{0}T+255(1-T)$, where $B_0$ is the original B value of the pixel and $T$ is the sky transmission calculated using Equation~\ref{sky_opacity}. A scale added to the top left part of the transmission map presents the colours of several levels of opacity such that $\Delta M = -2.5\log_{10}T$. Currently, opacity maps are evaluated visually by human observers, but the data may also be used to improve the effectiveness of robotic telescopes and autonomous sky surveys. Figure~\ref{opacity_map} is an example of an all-sky opacity map generated by the all-sky CONCAM camera operating in Cerro Pachon, Chile.
 A detailed description of the opacity maps is available at \citep{Sha05b}.

\subsection{Image compression}
\label{image_compression}

Some of the best sites for astronomical observations are located far from urban centers or other sources of light pollution. Therefore, a fast internet connection is not always available at some of these sites. Since FITS files consume a relatively large space, a data compression algorithm is sometimes required in order to maintain a high image-rate pipeline.

The FITS image compression algorithm used in the design described in this paper is {\it PHOTZIP} \citep{Sha05d}. PHOTZIP works by modeling, smoothing, and then compressing the astronomical background behind self-detected point sources while completely preserving values in and around those sources. Since the compression of the many background pixels is done in a symmetric fashion, the statistical properties such as mean and median (assuming normal distribution) are practically preserved. Since the values of the bright pixels are also preserved, photometry measurements of $I_{\mathrm star}-I_{\mathrm background}$ in the compressed/decompressed FITS images are practically identical to the original image.

Another advantage of PHOTZIP is that it allows users to control the compression/lossiness trade-off in terms of $\sigma$, so the compression factor can be adjusted to the available bandwidth at the site while users have an accurate sense of the lossiness due to the algorithm.

As an example of an existing application, the above algorithm is used regularly in CONCAM \citep{Nem99} all-sky camera operating at SALT observatory (South Africa) with a compression factor of 85\%. Experiments show that the photometric measurements in the compressed/decompressed images are practically the same as in the original images.

A more detailed description of the {\it PHOTZIP} algorithm is available at \citep{Sha05d}.

\subsection{Transient detection}
\label{transient_detection}

All-sky wide angle astronomical pipelines can be used for the purpose of optical transient detection \citep{Kro01}. Since the ability of all-sky optics to observe faint sources is not comparable to even simple narrow-angle telescopes, only bright transients can be detected by all-sky super-wide angle devices. However, the low limiting magnitude of all-sky optics can be weighed against the large portion of the sky covered by the wide lens. 

Brighter astronomical transients are sometimes considered more interesting to science \citep{Lin87,Men87}. Naturally, a wide-angle system is much more likely to capture such transients than narrow-angle telescopes covering only a very small portion of the sky at any given moment. In fact, one can reasonably assume that some bright short timescale transients visible to the unarmed eye are not noticed at all by the available narrow-angle sky surveys \citep{Sch05}.

In the proposed system, all-sky transient detection \citep{Sha05c2,Sha05e} requires several logical steps. The first is the rejection of pixels dominated by cosmic-ray generated counts, as described in Section~\ref{cosmic_ray_hits}. Next, bright planets and high-amplitude variable stars are also rejected using a star recognition algorithm designed to find astronomical objects in wide angle frames \citep{Sha05a}. After cosmic-ray hits and bright planets are rejected, the system searches for PSFs that are a given $\sigma$ brighter than the local background. Each PSF is then compared with the corresponding PSF in the canonical frame taken from the database of canonical images described in Section~\ref{opacity_maps}. The absence of moving parts in all-sky instruments simplifies the comparison of all-sky images so that finding corresponding PSFs in two images taken at the same sidereal time is relatively easy. The detected transients are listed in an xml-tagged HTML file and transmitted to the main server where they are available in real time. The current version of the transient detection mechanism does not provide information regarding the shape of the transient PSF so that users who wish to analyze the nature of the transient are required to observe the image by eye.

This system is currently operating in two all-sky cameras located in La Palma, Canary Islands and Cerro Pachon, Chile, and searches for optical transients brighter than 20$\sigma$ over their local background.

One down-side of this transient detection mechanism is noise produced by satellite glints \citep{Sch87}. This problem can be solved by comparing the detected transients with a database of predicted known satellites glints such as the one provided by {\it Heavens-Above.com}. This feature is not yet implemented in the current system. Experiments based on a device with a limiting stellar magnitude of 6.8 showed that approximately 5 bright transients are reported each night, most of which are satellite flashes and some others are bright meteors. Procedures performed in the server-side of the system reject satellite glints by comparing data from different cameras and by looking for sources that seem to rotate with the sky, as will be described in Section~\ref{server_transient_detection}.

\section{Server side software design}
\label{server_side}

The purpose of the server is to receive and archive the data sent from the nodes, generate GIF animations presenting the whole night as a movie, and provide web access to the data using a simple internet browser.

\subsection{Web server}
\label{web_server}

In order to increase the availability of the data, all files are stored on a web server and are made accessible via the internet. The web site is organized such that the front page is a world-map labeled with the locations of the active all-sky cameras. Clicking on a location leads to a web page showing the last image taken at that site. Each location has an archive that provides access to previous data. The archive provides the FITS images, opacity maps and annotated JPG images described in Section~\ref{star_recognition}. In addition, photometry files (also described in Section~\ref{star_recognition}) are also provided. The files are xml tagged, so that external software tools can easily collect data from the server. The URL of each file is also of a standard format which is {\it /xx/xxyymmdd/xxyymmddUThhnnss.html}, where {\it xx} is the symbol of the observatory, and {\it yy, mm, dd, hh, nn, ss} are the year, month, day, hour, minute and second of the beginning of the exposure, respectively. This file naming policy makes it simpler to find specific web pages, and also makes it easier for external automatic tools to navigate through the data.

The data are also copied to a database which allows performing simple queries. The database is based on the open source {\it My-SQL}. All queries are performed on-line using a web-based interface that supports queries according to stars (Henry Draper catalogue number), date and time.

Once the last image for the night is transmitted to the server, a GIF movie, which is a sequence of all images recorded that night, is generated on the server and placed in the archive. We found this feature useful for quick viewing of entire nights.

In addition, the web server also provides a discussion board where users can share their thoughts and experiences. The discussion board also serves as a log by documenting interesting events in a public fashion. 

\subsection{Transient detection (server side)}
\label{server_transient_detection}

In Section~\ref{transient_detection} we described a mechanism that detects transients in single images. The result of the single-image transient detection generated in the remote sites is transmitted to the server in the form of xml-tagged files. Each file contains a list of transients detected so far that night, and each entry in the list includes the date, time, approximate celestial coordinates (RA,DEC) and the filename of the image in which the transient was detected. The RA, DEC coordinates are obtained by using the fuzzy-logic based coordinate transformation algorithm described in \citep{Sha05a}.

A computer program running on the server reads the files transmitted by the remote stations and searches for transients detected in the same celestial coordinates and at the same time. A transient recorded by two separate all-sky cameras in the same region of the sky at the same time may indicate that the source is astronomical, and not a satellite or a bright meteor.

Another criterion taken into account is persistency. The system searches for transients appearing at the same geocentric celestial coordinates for at least two consecutive exposures taken by the same camera. The purpose of this mechanism is to find optical transients that rotate with the sky.

\subsection{Meteor science}

Networks of all-sky cameras are often used for meteor science \citep{Sop92,Obe98,Bro02,Spu03} and can provide a comprehensive coverage of meteor showers \citep{Raf01,Fle02,Bro02}. It also proved to be efficient in the search for sporadic fireballs \citep{Obe98,Spu03}. The described software system allows users to access images recorded during meteor showers and automatically detects sporadic bright fireballs as discussed in Section~\ref{transient_detection}.

Another feature related to meteor science using all-sky cameras is the 3D analysis of meteor trails by using data from two (or more) neighbouring all-sky cameras \citep{Sha05b2}. Since exposures at different stations start at the same time, it is likely that a meteor recorded by one camera is also recorded by another nearby camera. After transforming the image coordinates of the start and end of the meteor trail to celestial coordinates \citep{Sha05a}, simple triangulation can provide the altitude of the start of the meteor trail, the altitude of the end of the meteor trail, and the absolute trail's length with accuracy of $\sim$3\% \citep{Sha05b2}. When a meteor has more than two peaks of maximum brightness, the absolute distance between each peak can also be obtained \citep{Sha05b2}.

\section{Conclusions}

We described a software design that aims to provide a comprehensive software solution to all-sky panoramic astronomical pipelines. The software controls both the remote station, at which images are taken and processed, and the main server that archives and provides access to the data. The described software is currently used for the processing of all-sky panoramic astronomical pipelines generated by 11 all-sky cameras located in Mauna Kea and Haleakala (Hawaii), Cerro Pachon (Chile), Kitt Peak (Arizona), Mt. Wilson (California), Rosemary Hill (Florida), Siding Spring (Australia), Wise Obs. (Israel), Canary Islands (Spain), SALT (South Africa) and Hanle (India). The data collected in these observatories are available at http://nightskylive.net.

\label{last_page}

\end{document}